\documentclass[sn-jnl,referee]{sn-jnl}

\usepackage{graphicx}%
\usepackage{multirow}%
\usepackage{amsmath,amssymb,amsfonts}%
\usepackage{amsthm}%
\usepackage{mathrsfs}%
\usepackage[title]{appendix}%
\usepackage{xcolor}%
\usepackage{textcomp}%
\usepackage{manyfoot}
\usepackage{booktabs}%
\usepackage{ragged2e}
\usepackage{algorithmicx}%
\usepackage{algpseudocode}%
\usepackage{listings}%
\usepackage{tabularx}%

\usepackage{pdflscape}
\usepackage{array}
\usepackage{multirow}
\usepackage{booktabs}
\usepackage{enumitem}
\usepackage{xcolor}

\usepackage{makecell}
\usepackage{lscape}
\usepackage{adjustbox}
\usepackage[normalem]{ulem}
\usepackage{hyperref}
\newcommand{\siref}[2]{\hyperref[#2]{#1}}
\usepackage{pdflscape} 

\usepackage[numbers]{natbib}
\usepackage[resetlabels]{multibib}
\newcites{SI}{References}
\usepackage{graphicx}
\usepackage{lineno}

\graphicspath{{figures/}}

\theoremstyle{thmstyleone}%
\theoremstyle{thmstyletwo}%

\theoremstyle{thmstylethree}%

\begin{document}

\title[Article Title]{The Uneven Decline of Collective Knowledge Production: Evidence from Stack Overflow After Generative AI}


\author[1]{\fnm{Myokyung} \sur{Han}}

\author[1]{\fnm{Taegyoon} \sur{Kim}}

\author*[2]{\fnm{Jinhyuk} \sur{Yun}}
\email{jinhyuk.yun@ssu.ac.kr}

\author*[1]{\fnm{Lanu} \sur{Kim}}
\email{lanukim@kaist.ac.kr}

\affil[1]{
\orgdiv{School of Digital Humanities and Computational Social Sciences},
\orgname{KAIST},
\orgaddress{
\street{291 Daehak-ro, Yuseong-gu},
\city{Daejeon},
\postcode{34141},
\country{Republic of Korea}
}
}

\affil[2]{
\orgdiv{School of AI Convergence},
\orgname{Soongsil University},
\orgaddress{
\street{369 Sangdo-ro, Dongjak-gu},
\city{Seoul},
\postcode{06978},
\country{Republic of Korea}
}
}

\abstract{Generative AI (Gen AI) is reshaping how individuals learn and work, but its consequences for collective knowledge, the shared body of knowledge that online communities produce together, remain poorly understood. Prior work has documented an aggregate decline in participation on knowledge-sharing platforms, but it remains unclear which specific kinds of knowledge are being lost first. We study this question using Stack Overflow, one of the largest online communities for software engineering, treating the release of ChatGPT-3.5 as a natural shock. Analyzing over two million questions posted between 2020 and 2025, we track how two dimensions of collective knowledge, difficulty and data availability, change following Gen AI's release. Using diverse methods and robust checks, we find consistent patterns. Easy questions decline sharply while difficult questions become more common, a pattern corroborated by rising code complexity. Data-rich topics and tags lose share of questions, while data-scarce ones gain ground. The two dimensions also interact: the decline in easy questions is concentrated specifically within data-rich domains, while difficult questions increase regardless of data availability. This pattern extends beyond Python across programming languages, with more prevalent languages showing sharper shifts. Together, our findings reveal that Gen AI's impact on collective knowledge is uneven, eroding easy, accessible knowledge first while more complex, less common knowledge persists.}

\keywords{\textnormal{LLM}, \textnormal{Automation}, \textnormal{Software Engineering}, \textnormal{Stack Overflow}, \textnormal{Task-based approach}}

\maketitle

\section{Introduction}\label{sec1}


Generative AI (Gen AI) is changing how humans learn, and concern about its cognitive costs is increasing from multiple directions. A growing body of academic research documents how reliance on Gen AI may undermine effortful cognitive engagement: although Gen AI allows people to obtain answers quickly, this convenience may come at the expense of the deep thinking and processing that learning requires\citep{kosmyna2025your, ke2026ai}. This concern is not confined to academic circles. Educators and practitioners at the frontlines of teaching have increasingly observed similar patterns among students and voiced them in public discourse \citep{ipsos2026teachers, roytburg2026students}. Taken together, these accounts point to a shared concern: Gen AI is already beginning to reshape individuals' cognitive processes.

Although concern about what individuals are losing through their use of Gen AI continues to grow, comparatively little attention has been paid to what we might be losing as a collective. After all, knowledge is not produced by individuals alone. Recent work has begun to address this question in the context of scientific knowledge, showing that Gen AI is already reshaping how scientific knowledge is produced \citep{hao2026artificial}. However, scientific knowledge represents just one form of collective knowledge, capturing cutting-edge innovation. Collective knowledge also takes a far more practical, everyday form—the kind people rely on to solve real problems \citep{malone2010collective, bernstein2018intermittent}. Such knowledge is built through people's long-standing, voluntary willingness to contribute to the construction of shared knowledge \citep{ayoubi2023knowledge, burton2024large}. This form of contribution has traditionally been channeled through public repositories such as Wikipedia and Stack Overflow, which have accumulated vast bodies of human knowledge over time \citep{mesgari2015sum, barua2014developers}. Whether and how Gen AI is reshaping this practical, everyday form of collective knowledge remains largely unknown: what kind of knowledge, as a collective, are we losing first?

So far, scholars have documented declining participation and user engagement across several public knowledge-sharing platforms, such as Wikipedia and Stack Overflow, following Gen AI's release \citep{del2024large, burtch2024consequences, lyu2025wikipedia}. Notably, this decline does not appear to reflect a general reduction in people's engagement with online communities: platforms centered on social interaction, such as Reddit, have not seen a comparable decline over the same period \citep{burtch2024consequences}. This contrast suggests that what Gen AI is displacing is not online participation itself, but specifically the collective production of knowledge. Yet these aggregate trends capture only the overall magnitude of this decline; they do not tell us which kinds of knowledge are being lost. This gap points to our analytic focus: the very questions that document this aggregate decline can also serve as a lens for identifying which specific kinds of knowledge we are collectively failing to produce.

We propose two hypotheses about what kind of knowledge is being lost. First, we hypothesize that the knowledge that has stopped being produced is disproportionately the kind that is cognitively easy. Gen AI has proven particularly powerful at cognitive, text-based tasks, and its capabilities have advanced considerably in solving increasingly complex problems over time. Historically, progress in AI has often been marked by its ability to solve tasks once thought to require high-level cognition—first chess \citep{campbell2002deep}, then Go \citep{silverMasteringGameGo2016}, and more recently, tasks such as competitive programming \citep{li2022competition}, mathematical olympiad problem-solving \citep{trinh2024solving}, and scientific discovery, including protein structure prediction \citep{jumper2021highly}. This trajectory suggests that Gen AI is continually pushing the boundary of the cognitively demanding work it can perform, making it increasingly capable of resolving easier questions first.

Second, we hypothesize that human knowledge production declines more sharply in areas where Gen AI is most effective due to high training data availability. This follows from the fact that Gen AI is fundamentally a language model trained on existing data, and, consistent with the scaling law, it performs better when trained on larger corpora \citep{kaplan2020scaling}. Indeed, LLMs have been shown to underperform in several low-resource domains, precisely because insufficient training data exists for these areas. This includes rare or niche programming languages that are underrepresented in training corpora \citep{joel2024survey}, as well as low-resource human languages, which show substantially weaker performance compared to high-resource ones such as English \citep{joshi2020state, lai2023chatgpt, huang2026survey}.

We examine changes in collective knowledge through the lens of two dimensions: difficulty and data availability. To study this question, we focus on Stack Overflow, one of the largest online knowledge-sharing communities dedicated to software engineering. Stack Overflow is particularly well-suited to this question for three reasons. First, software engineering is among the fields most rapidly and directly affected by Gen AI \citep{daniotti2026using}, reshaping employment opportunities \citep{brynjolfsson2025canaries} and shifting student interest in computer science \citep{ovide2026hottest}. Moreover, unlike licensed professions such as medicine or law, it lacks institutional buffers against this disruption. This makes it an ideal setting for observing how Gen AI reshapes collective knowledge. Second, Stack Overflow has operated continuously since 2008, providing consistent data both before and after Gen AI's release in November 2022 and allowing us to treat this release as a natural shock. Third, Stack Overflow banned Gen AI-generated content shortly after ChatGPT-3.5's release \citep{SOban2024}. Although imperfectly enforced, this policy allows us to treat remaining post-shock content as a proxy for genuine human production.


For the analysis, we treat ChatGPT-3.5's release date as the key reference point for our analytic window, studying Stack Overflow questions posted over a five-year period spanning two years before to three years after this date (2020–2025). Within this window, we examine how human-generated questions have changed before and after Gen AI's release along two dimensions: difficulty and data availability. Because both dimensions are abstract constructs, we measure each using two complementary operationalizations—one grounded in human judgment, the other in machine-based metrics—allowing us to assess the robustness of our findings across independent measures. We first trace how each dimension changes over time within Python-related questions, then examine how difficulty and data availability jointly evolve, and finally test whether these patterns extend beyond Python to a broader set of programming languages.

To measure difficulty, we analyze the content of each question, both its natural-language description and its embedded source code, to assess how cognitively demanding it is. To measure data availability, we trace how much data had accumulated on Stack Overflow prior to Gen AI's release, using this as a proxy for the knowledge density available to Gen AI models at the time. This proxy is not without grounding: just before frontier AI developers transitioned to closed-data regimes, the last generation of transparent foundation models, such as LLaMA~\citep{touvron2023llama} and those trained on The Pile~\citep{gao2020pile}, explicitly incorporated Stack Exchange data as a core component. This indicates that around the release of ChatGPT-3.5, our study's critical intervention point, Gen AI models were heavily reliant on this repository for their programming capabilities. Underscoring the continued relevance of this data, OpenAI later established an official partnership with Stack Overflow in 2024 to directly integrate the platform's knowledge into GPT-models \citep{stackoverflow2024openai}. Because LLM performance is governed by the scaling law \citep{kaplan2020scaling}, the volume of accumulated Stack Overflow questions on a given topic offers a reasonable, if indirect, indicator of how effectively Gen AI can address problems in that domain.

We identify four consistent findings. First, question difficulty has increased overall—easy (basic and intermediate) questions have declined while difficult (advanced) questions have become more common, consistent with rising code complexity. Second, data-rich topics and tags have declined in their share of all questions, while data-scarce ones have gained share. Third, when considering difficulty and data availability jointly, we find that the decline in easy questions is concentrated almost entirely within data-rich topics, whereas difficult questions rose regardless of data richness. Fourth, these patterns extend beyond Python: across 30 programming languages, difficulty increases and data availability decreases overall, with the magnitude of change proportional to each language's pre-Gen AI data richness—more prevalent languages exhibit sharper shifts, while less common languages change comparatively little.

Taken together, our findings suggest that Gen AI's reshaping of collective knowledge has been uneven, concentrated on easier and more data-rich domains while sparing more complex, less common knowledge. Using software engineering as a case study, our findings speak to a broader question: not simply whether collective knowledge is diminishing, but where it is likely to erode first. By jointly analyzing difficulty and data availability, two dimensions rooted in how large language models are built and trained, our framework offers a systematic way to identify which parts of humanity's collectively produced knowledge are most vulnerable as Gen AI continues to advance. 

\section{Results}\label{sec2}

\subsection{Changes in Difficulty}\label{subsec1}

\begin{figure}[!t]
\centering
\includegraphics[width=\linewidth]{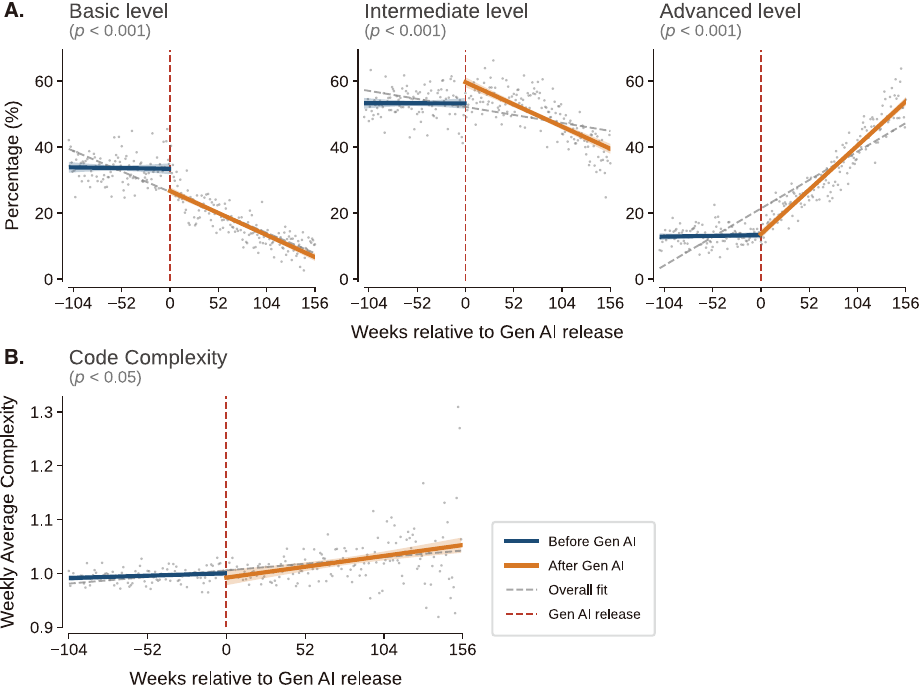}
\caption{Changes in question difficulty and code complexity. (A) Weekly proportion of questions classified as Basic, Intermediate, or Advanced, based on our AI annotator (see \siref{Supplementary Information}{sec:FE_2.A}). We randomly sample 30 questions per day over the 260-week (1,820-day) study period, yielding 54,600 classified questions in total (see \siref{Supplementary Information}{sec:FE_2.A.8} for results using alternative sampling strategies). Chow tests indicate significant slope changes following Gen AI's release for all three difficulty levels (Basic: $F = 49.662$; Intermediate: $F = 68.067$; Advanced: $F = 224.326$; all $p < 0.05$). (B) Weekly average cyclomatic complexity of source code embedded in questions. A Chow test indicates a significant slope change following Gen AI's release ($F = 4.145$, $p < 0.05$). In both panels, the x-axis denotes the number of weeks relative to ChatGPT-3.5's release on November 30, 2022 (week 0); the y-axis denotes the percentage of questions at each difficulty level (A) or the average cyclomatic complexity score (B).}\label{C_Result_Fig1}
\end{figure}

Our first research question examines whether the difficulty of Stack Overflow questions has changed since the release of Gen AI, from two perspectives: human cognition (Fig.\ref{C_Result_Fig1}A) and machine assessment (Fig.\ref{C_Result_Fig1}B). For the human-perspective measure, we track the weekly percentage of Stack Overflow questions across three difficulty levels: Basic, Intermediate, and Advanced. As shown in Fig.\ref{C_Result_Fig1}A, the percentage of intermediate-level questions, which dominated before Gen AI's release (average 53.31\%), decreases to 35.08\% in week 156. Basic-level questions represent the second-largest category before Gen AI's release (average 33.49\%) but decline steadily afterward, eventually comprising only about 6.7\% of all questions and becoming the smallest category 156 weeks after the release. In contrast, advanced-level questions increase continuously over the same period, rising from 15\% to 53.8\%—meaning that more than half of all questions are now considered difficult for users to solve. Chow test results indicate that the slope changes after Gen AI's introduction are statistically significant for all three difficulty levels (basic, intermediate, and advanced) at the 0.05 alpha level. 

In Fig.\ref{C_Result_Fig1}B, we test the same idea using an alternative measure of difficulty—source code complexity—which captures the complexity of code embedded in each question, serving as a proxy for machine-assessed difficulty. The weekly average code complexity remains stable during the year preceding Gen AI's release but begins to increase steadily afterward. According to the Chow test, this change in slope is statistically significant at the 0.05 alpha level, indicating that questions now involve more complex programming challenges than before.

Although these two measures rely on very different approaches, they point in the same direction: questions posted on Stack Overflow have become more difficult since Gen AI's release. This evidence suggests that users no longer see the need to post simple problems on the website, as Gen AI can now fulfill that role. This trend has continued to intensify in three years since.

\subsection{Changes in Data Availability}\label{subsec2}

We now examine the second characteristic: whether the size of accumulated data facilitates the use of Gen AI. To test this, we compare the weekly distribution of data-rich (top 20\%) and data-scarce (bottom 20\%) topics over time. The top and bottom 20\% topics are defined based on their frequency in the pre-Gen AI period; we then track how these previously popular and rare topics evolve after Gen AI's release. We assign topics using two approaches: an unsupervised machine learning model, BERTopic (see \siref{Supplementary Information}{sec:FE_2.C.3} for model specification), and tags attached by users themselves. These two approaches are based on similar ideas, categorizing a question into a certain topic, but differ in the granularity of topics. For example, the top 20\% of topics range from simple built-in function topics (e.g., loop, regex, numpy) to web scraping topics (e.g., beautifulsoup, scrape), while the bottom 20\% include specialized topics such as testing-related topics (e.g., pytest, yaml, mock), anti-blocking scraping topics (e.g., tweets, youtube, proxy), and framework-based topics (e.g., mongodb, pyspark, kafka, databricks). Similarly, the top 20\% of tags include \texttt{<numpy-slicing>} , \texttt{<infinite-loop>}, and \texttt{<beautifulsoup>}, while the bottom 20\% include \texttt{<pytest-selenium>}, \texttt{<proxy-authentication>}, and \texttt{<mongodb-geospatial>}; tags, however, have a narrower scope than topics. 

The two approaches also differ in how evenly they distribute questions across topics. BERTopic assigns topics in relatively even proportions: at the beginning of the observed period, the top 20\% of topics account for about 0.35 of all questions, while the bottom 20\% account for around 0.08. User-generated tags, in contrast, are far more concentrated. The top 20\% of tags account for nearly all tag assignments, while the bottom 20\% make up only about 0.002 of the total. This concentration likely arises from a key difference between the two approaches: BERTopic applies consistent criteria to define topic scope and proportion, whereas tag assignment depends on individual user behavior. In particular, users tend to attach popular tags to their questions to increase visibility in search results.

\begin{figure*}[!t]
\centering
\includegraphics[width=.8\textwidth]{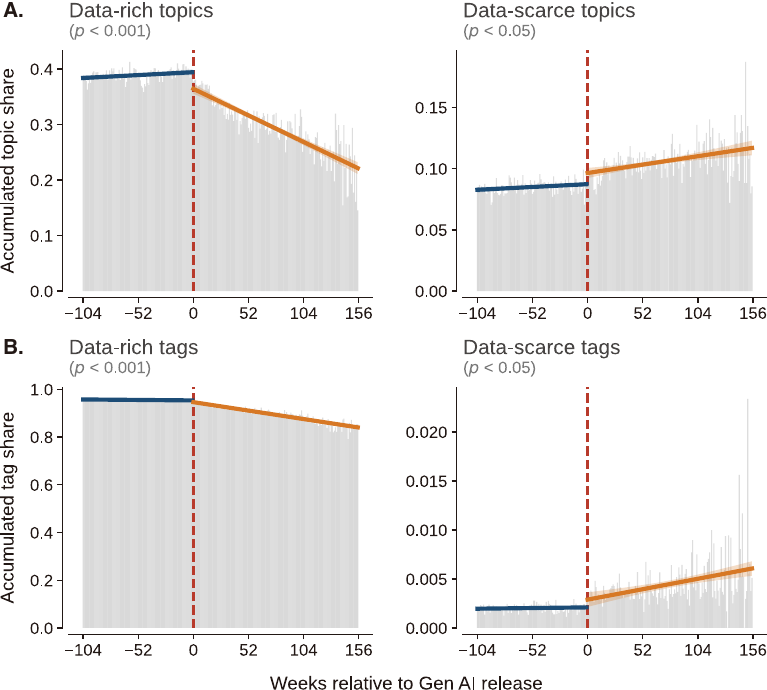}
\caption{Changes in data availability. (A) Weekly composition of the data-rich(top 20\%, $N = 10$) and data-scarce(bottom 20\%, $N = 10$) machine-assigned topics, identified via BERTopic, before and after Gen AI's release. We identify the data-rich and data-scarce topics by frequency in the pre-Gen AI period and visualize their weekly composition using stacked bar graphs. The x-axis denotes the number of weeks relative to Gen AI's release (week 0); the y-axis denotes the cumulative share of questions attributable to each group of topics. (B) The same analysis applied to user-assigned tags, comparing the top 20\% ($N = 3{,}255$) and bottom 20\% ($N = 3{,}255$) of tags by frequency. For both panels, Chow tests indicate significant slope changes following Gen AI's release (topics: $F = 92.297$ for top 20\%, $F = 6.356$ for bottom 20\%; tags: $F = 155.021$ for top 20\%, $F = 4.236$ for bottom 20\%; all $p < 0.05$). All the top and bottom 20\% are defined based on their frequency in the pre-Gen AI period.}
\label{C_Result_Fig2}
\end{figure*}

We next examine how topic composition changes between the pre- and post-Gen AI periods, analyzing machine-assigned topics through descriptive visualization of both data-rich and data-scarce topics. As shown in Fig.\ref{C_Result_Fig2}A, we observe a decline in the proportion of data-rich topics, accompanied by an increase in the proportion of data-scarce topics. This shift suggests that data-rich knowledge areas may be more easily addressed by Gen AI, whereas data-scarce, niche areas continue to require human software engineering expertise. Overall, this indicates that topic composition has become more balanced compared with the pre-Gen AI period, as commonly discussed subjects decline while rare topics gain prominence.

For robustness, we conducted two additional checks. First, we replaced BERTopic with LDA and found similar patterns (see \siref{Supplementary Information}{sec:FE_2.C.5}). Second, we repeated the same analysis using an earlier, comparable time window (2019–2022) to rule out the possibility that the same pattern would have emerged even without the introduction of Gen AI. The results show no similar distributional shift during this earlier window, supporting the interpretation that the observed pattern is specific to the post-Gen AI period (see \siref{Supplementary Information}{sec:FE_2.C.6}).

We also examine distributional shifts using tag information, applying a parallel set of analyses to those conducted with topic data (Fig.\ref{C_Result_Fig2}B). Although topics and tags represent two different operationalizations of the same underlying concept—grouping questions by subject matter—the overall trend remains strikingly similar between the two: the share of the top 20\% decreases, while the share of the bottom 20\% increases after Gen AI's release. This convergence across two distinct measurement approaches suggests that the observed pattern is robust.

\subsection{Interaction between Difficulty and Data Availability}\label{subsec3}
\begin{figure*}[t]  
\centering
\includegraphics[width=\textwidth]{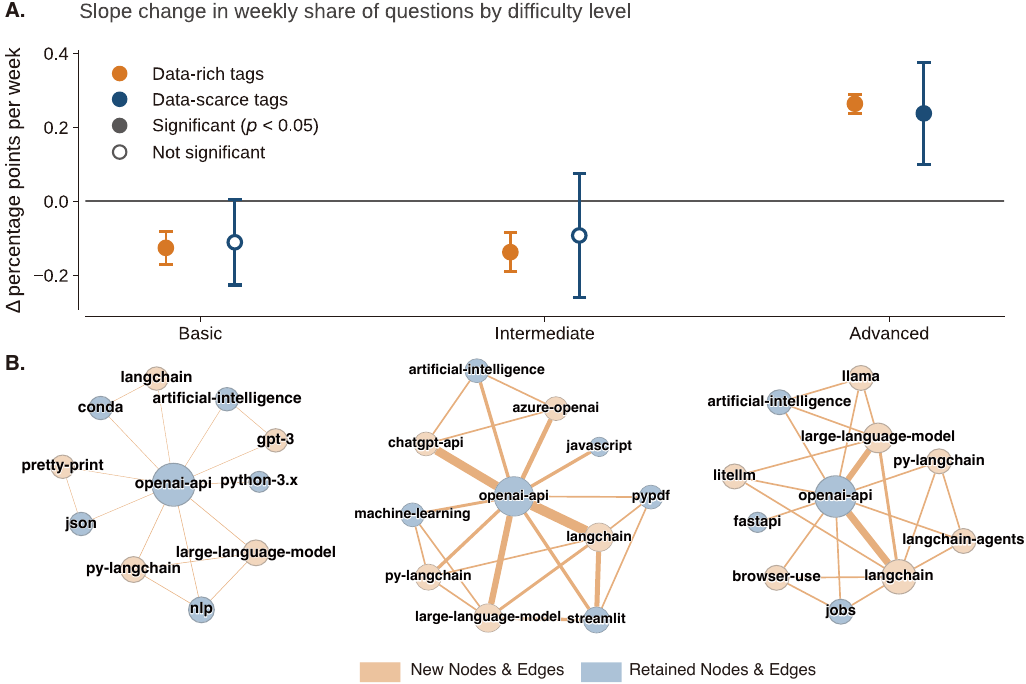}
\caption{Interaction between difficulty and data availability. (A) Weekly proportion of questions at each difficulty level (Basic, Intermediate, Advanced), compared between data-rich (top 20\% tags) and data-scarce (bottom 20\% tags) questions. We estimate an interrupted time-series model for each difficulty level and tag group, testing whether the weekly proportion trend changes significantly following Gen AI's release ($\alpha = 0.05$; see \siref{Supplementary Information}{sec:FE_3.A.2} for full estimates). (B) As an illustrative example, we construct co-occurrence tag networks centered on \texttt{<openai-api>}, a data-scarce tag whose accumulated share increased markedly after Gen AI's release (see main text for details), separately for Basic- (left), Intermediate- (middle), and Advanced-level (right) questions. Each node represents a tag, and edges connect tags that co-occur within the same question (see \siref{Supplementary Information}{sec:FE_3.A.3}).}

\label{C_Result_Fig3}
\end{figure*}

So far, we have analyzed patterns of difficulty and data availability independently. We now examine how these two factors co-evolve by tracing the weekly proportion of questions across difficulty levels (Basic, Intermediate, Advanced), separately for data-rich (top 20\% tags) and data-scarce (bottom 20\% tags) questions. For each difficulty level, we estimate the change in weekly trend following Gen AI's release using an interrupted time-series (ITS) model (see \siref{Supplementary Information}{sec:FE_3.A.2}). The ITS model tests whether an intervention—here, the release of Gen AI—produces a significant change in the level and/or slope of an outcome variable relative to its pre-intervention trend, allowing us to distinguish changes associated with Gen AI's release from trends that were already underway beforehand \citep{bernal2017interrupted}.

As shown in Fig.\ref{C_Result_Fig3}A, questions associated with data-rich tags exhibit distinct patterns across difficulty levels. Consistent with the overall trend in Fig.\ref{C_Result_Fig1}A, the weekly proportion of basic- and intermediate-level questions declines at a significantly steeper rate after Gen AI's release (by approximately 0.13 and 0.14 percentage points per week, respectively), while the proportion of advanced-level questions increases at a significantly steeper rate (by 0.26 percentage points per week). This pattern suggests that the overall decline in data-rich tags' share (Fig.\ref{C_Result_Fig2}B) is driven primarily by fewer basic- and intermediate-level questions, rather than by advanced-level questions, within this category. Data-scarce tags, by contrast, show weak, non-significant changes at the basic and intermediate levels (-0.11 and -0.09 percentage points per week, respectively); only the advanced level shows a statistically significant change (0.24 percentage points per week), suggesting that the growing share of data-scarce questions is driven primarily by advanced-level questions.

As an illustrative example of the observed pattern, we select \texttt{<openai-api>}, a data-scarce tag that nonetheless shows a dramatic increase in accumulated share relative to the pre-Gen AI period (approximately 18.75-fold, from 0.008\% to 0.150\%). Centered on this tag, we construct a co-occurrence egocentric network covering all time periods in our analysis, and show its formation by basic-, intermediate-, advanced-level questions (Fig.\ref{C_Result_Fig3}B). The node colors indicate each associated tag's category relative to the pre-Gen AI period: new (first appearing after Gen AI's release) and retained (present in both periods). For instance, in Panel B, tags such as \texttt{<nlp>} and \texttt{<conda>} appear in both periods and are therefore classified as retained, while tags such as \texttt{<large-language-model>} and \texttt{<py-langchain>} first appear only after Gen AI's release and are therefore classified as new. To ensure comparability across difficulty levels, we extract the top 10 nodes by eigenvector centrality for each network in Panel B (see \siref{Supplementary Information}{sec:FE_3.A.3}).

Fig.\ref{C_Result_Fig3}B reveals broadly similar yet technically distinct application patterns across difficulty levels. All three levels share common tags such as \texttt{<langchain>} and \texttt{<large-language-model>}, reflecting their close technical relationship with \texttt{<openai-api>}. However, the overall composition of nodes differs considerably across levels: 50\% of nodes adjacent to \texttt{<openai-api>} are retained for basic- and intermediate-level questions, compared with only 30\% for advanced-level questions, suggesting that advanced-level questions engage more actively with newly emerged topics. In addition, the density of the ego network is higher for intermediate- and advanced-level questions (0.40) than for basic-level questions (0.29), indicating that tags surrounding \texttt{<openai-api>} are more densely connected to each other at higher difficulty levels.

More specifically, basic-level questions are associated with fundamental tags related to file input/output (e.g., \texttt{<json>}, \texttt{<pretty-print>}) or environment setup (e.g., \texttt{<conda>}). Advanced-level questions, by contrast, are connected to framework-level tags: alongside the common tags \texttt{<langchain>} and \texttt{<large-language-model>}, numerous new tags emerge, such as \texttt{<litellm>} and \texttt{<langchain-agent>}. Intermediate-level questions occupy a middle ground: they connect to some framework-level tags as well, but are more strongly linked to retained tags, consistent with the higher retention rate reported above (see \siref{Supplementary Information}{sec:FE_3.A.3}).

\subsection{Heterogeneous patterns across programming languages}\label{subsec4}  
\begin{figure*}[!t]  
\centering
\includegraphics[width=.9\textwidth]{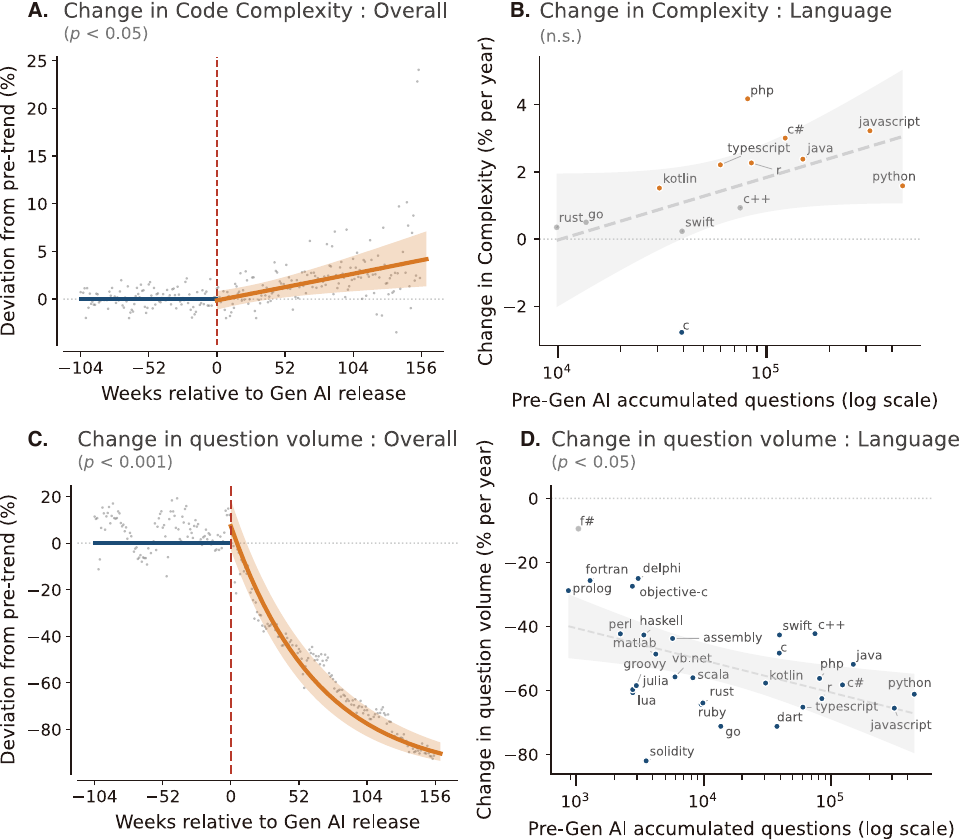}
\caption{Heterogeneous patterns across programming languages. (A) Estimated deviation in cyclomatic complexity from the pre-Gen AI trend, aggregated across the 13 programming languages for which complexity scores could be computed, using an interrupted time-series (ITS) model. (B) Language-specific slope changes in complexity, obtained by applying the same ITS model separately to each language. Each dot represents one language: the x-axis shows the language's accumulated question volume prior to Gen AI's release, and the y-axis shows the yearly change in the slope of complexity following its release. Dot color indicates the statistical significance of each language's slope change: red denotes a significant positive change, blue a significant negative change, and gray a non-significant change ($\alpha = 0.05$). An ordinary least squares (OLS) regression indicates a positive, though not statistically significant, relationship between pre-Gen AI question volume and the magnitude of this slope change ($R^2 = 0.26$, $\text{slope} = 0.81$). (C) Estimated deviation in question volume from the pre-Gen AI trend, aggregated across all 30 programming languages, using the same ITS model. (D) Language-specific slope changes in question volume, obtained by applying the same ITS model separately to each of the 30 languages. As in (B), the x-axis shows each language's accumulated pre-Gen AI question volume, the y-axis shows the yearly change in the slope of question volume following Gen AI's release, and dot colors follow the same convention as in (B). An OLS regression indicates a negative and statistically significant relationship between pre-Gen AI question volume and the magnitude of this slope change ($R^2 = 0.23$, $\text{slope} = -4.36$) (see \siref{Supplementary Information}{sec:FE_3.B}).}
\label{C_Result_Fig4}
\end{figure*}

As the final analysis, we broaden our scope beyond Python to examine other programming languages. This extension is motivated by an earlier finding: the share of data-rich topics shifted substantially in the post-Gen AI period, yet this pattern was identified using Python-related questions alone—the most data-rich programming language on Stack Overflow. This raises the question of whether the dynamics observed within Python also hold when aggregated across programming languages. To address this, we apply the same framework to this extended scope. We present two pairs of panels: difficulty (Panels A and B) and data availability (Panels C and D). For each pair, we first plot the overall trend across programming languages (Panels A and C), and then examine whether this trend differs across individual languages (Panels B and D).

We first measure difficulty using code complexity.\footnote{We omit the human-coded difficulty measure here because calibrating equivalent difficulty levels across languages is not straightforward, and scaling this measure through additional human annotation was not feasible for this analysis.} By analyzing 13 languages with sufficient code snippets available to compute code complexity (i.e., languages with at least 30 code snippets per week for at least 50 weeks both before and after week 0, totaling over 100 qualifying weeks), the combined overall code complexity increased in the post-Gen AI period, rising by approximately 4.9\% relative to the counterfactual pre-trend by week 156 (the end of the study window). We also observe that the variance in complexity across individual questions grows over this period, indicating that some questions become substantially more complex than others even within the same week.

To examine whether this trend varies across languages, we estimate the yearly change in complexity for each language and relate it to the question volume accumulated in that language during the pre-Gen AI period(Panel B in Fig. \ref{C_Result_Fig4}). This analysis reveals a positive, though not statistically significant, relationship ($R^2 = 0.26$, $\text{slope} = 0.81$): languages with richer pre-Gen AI data—that is, a larger accumulated question volume—tend to exhibit larger increases in code complexity following Gen AI's release. By contrast, data-scarce languages (e.g., Rust, Go) show no significant change, and some (e.g., C) even show a decline in complexity.

As a second approach, we examine data availability across all 30 programming languages.\footnote{Unlike code complexity, this analysis can be extended to the full set of 30 languages, as sufficient question volume is available for each.} We apply the same framework to overall question volume, which declines significantly in the post-Gen AI period, falling by approximately 80\% relative to the counterfactual pre-trend by week 156  following its release. This rapid decline in community activity has also been documented in prior research \citep{burtch2024consequences, del2024large}. While all 30 languages decline over this period, the yearly rate varies considerably across languages, as shown in Panel D. Relating this decline to each language's pre-Gen AI question volume, we find a negative and statistically significant relationship ($R^2 = 0.23$, $\text{slope} = -4.36$): programming languages that were more data-rich in the pre-Gen AI period tend to experience a stronger decline in question volume, whereas data-scarce languages exhibit a relatively weak decline in the post-Gen AI period.

\section{Discussion}\label{sec12}

So far, we have examined Stack Overflow questions using two key measures, difficulty and data availability, over a five-year period spanning before and after the release of ChatGPT-3.5, the first widely adopted commercial Gen AI model. We have also extended this analysis beyond Python to test whether these patterns generalize across programming languages. Across multiple measures, the trends were consistent. The share of advanced questions and overall code complexity increased following Gen AI's release, indicating a shift toward more complex problem-solving activity on the platform. Regarding data availability, questions tied to data-scarce topics became relatively more prevalent, while those tied to data-rich topics declined. When considered jointly, these two dimensions reveal that questions which were both easy and tied to data-rich topics were the most likely to disappear in the post-Gen AI period. Extending this analysis across 30 programming languages reveals a consistent pattern of uneven erosion in collective knowledge: more prevalent languages experienced sharper increases in complexity and sharper declines in data availability, while less common languages changed comparatively little.

This cross-language pattern is clearly illustrated by Python itself. Python, the most data-rich programming language \textcolor{blue}{in Stack Overflow}, with the largest volume of questions among all tags from 2020 to 2025, showed a relatively large decline in posts in the post-Gen AI period, compared to data-scarce languages such as Prolog and Fortran. A similar pattern was observed on Wikipedia: growth in views and edits stagnated, especially for content readily available to Gen AI \citep{lyu2025wikipedia, reeves2025exploring}, underscoring the importance of data availability. In terms of difficulty, data-rich languages such as Python also showed a marked increase in code complexity, whereas data-scarce languages did not exhibit a comparable rise. Given this uneven increase in code complexity depending on data availability, we infer that data availability is one of the key variables explaining the uneven erosion of collective knowledge.

Our findings also point to a possible polarization within data-rich areas, driven by the sharper rise in difficulty these areas experienced relative to data-scarce ones. Beginners in technical domains have traditionally relied on relatively easy problems as an entry point \citep{von2003community, steinmacher2018let}; our findings suggest that such entry points may be disappearing, particularly within data-rich domains. This pattern echoes findings in the automation literature, which suggests that junior engineers, who typically handle less complex tasks, are beginning to face declining job opportunities \citep{dell2023navigating, brynjolfsson2025canaries}.

This possibility is further supported by a supplementary analysis of salary data from Stack Overflow's annual Developer Survey (2019–2025), comparing changes in the salary premium across programming languages for junior- and senior-level developers before and after Gen AI's release (see \siref{Supplementary Information}{sec:FE_4.A}). As shown in Fig.\ref{SI_FE_4.A_Fig1}, the salary premium declined more sharply for junior-level developers in 11 of 17 languages. Because junior-level salary premiums in engineering have typically exceeded those of senior developers, reflecting the field's fast-changing nature \citep{deming2020earnings}, this decline suggests that Gen AI's effects have fallen disproportionately on less experienced developers. Together, these patterns suggest that within data-rich, prevalent domains, engineers with advanced, less automatable skills are likely to thrive, while those relying on basic or common skills may face diminishing opportunities—potentially reshaping the distribution of "good" jobs toward those with greater access to the human, economic, and social capital needed to cultivate rare, sophisticated expertise.

Taken more broadly, our findings point to a substantial narrowing of human-generated collective knowledge. As shown in the figure, both the overall volume of posts and the number of posts across programming languages declined in the post-Gen AI period, consistent with prior literature \citep{burtch2024consequences, quinn2025heterogeneous}. This narrowing continued to intensify over time when we extended the analytic window to include the period up to November 2025. One possible explanation is that Gen AI is undermining the very motivation that has historically sustained collective knowledge production. Voluntary contribution has long been fueled by the desire to help others solve problems they could not solve alone; as Gen AI increasingly performs this role on its own, that motivation may erode, raising the risk that collective knowledge as a whole could collapse \citep{acemoglu2026ai, peterson2025ai}.

So far, we have operationalized data availability as the volume of digitized text documenting software engineering knowledge. It is worth noting, however, that the boundary of what counts as "digitized" data is itself expanding. Recently, gig-economy platforms such as Instawork have begun connecting companies with members of the general public, who wear headband-mounted cameras while performing everyday household tasks, cooking, cleaning, watering plants, so that the footage can be used to train robots \citep{christopher2026strapping}. Physical movement data of this kind have historically been far scarcer than text, which is one reason physical labor occupations have remained comparatively insulated from automation relative to cognitive occupations \citep{freyFutureEmploymentHow2017, firoozi2025foundation}. If the same scaling-law logic underlying our findings extends to this new data modality, however, this pattern may reverse: as physical movement data accumulate, physical labor may begin to face the same data-driven exposure that software engineering questions have already experienced in our analysis.  

Our work has several limitations. First, the constraint to the software engineering domain would be one of the limitations. Even though software engineering well suited for analysis as the field had long-standing culture of knowledge sharing \citep{vasilescu2013stackoverflow, ghobadi2015drives}, it would still be hard to generalize our findings to other field of knowledge as a whole. Second, a limitation naturally follows from the first: we examined only a single community, Stack Overflow. Although we selected this community for its size and global accessibility, inherent selection biases are potentially present. Stack Overflow likely attracts engineers with specific traits, particularly those inclined toward knowledge sharing and collaborative problem-solving, which may limit the generalizability of our findings to the broader software engineering population. Even though these limitations still exist, our work provides crucial quantitative evidence of how collaborative knowledge has undergone rapid erosion driven by AI, which is otherwise difficult to observe directly. By capturing these shifts in real time through large-scale data, we offer measurable insights into AI's impact on the landscape of collaborative knowledge at a critical moment when such empirical study is most needed.

Future research could take several exciting directions. Beyond the difficulty and data availability of questions, one direction is to examine the interaction patterns underlying knowledge exchange itself. Using network methodology, future work could investigate how knowledge-sharing networks evolve over time in response to Gen AI. Moreover, Stack Overflow offers a unique and high-resolution window into individual learning trajectories. Its granular activity logs, spanning back to 2008, allow researchers to trace how users acquire and apply knowledge over time. Studying these patterns could reveal how software engineers develop expertise and provide a unique perspective on how AI tools may reshape, or even replace, traditional skill-building and knowledge accumulation processes in software engineering.

\section{Data and Methods}
\subsection{Data}
We construct our dataset from Stack Overflow's publicly available historical records, using the January 6, 2026 snapshot, which contains 60,371,716 questions and answers spanning Stack Overflow's launch on July 31, 2008, through January 6, 2026. From this 18-year dataset, we center our study on the initial release of ChatGPT 3.5 on November 30, 2022. ChatGPT 3.5 serves as a key reference point for two reasons: first, it marked the first widely adopted commercial Gen AI model \citep{NYTchatgpt}; and second, Stack Overflow imposed restrictions on AI-produced content immediately following its release \citep{SOban2024}. This restriction allows us to observe how human-driven knowledge changed in response to Gen AI's release.

Based on this reference point, we define a five-year analytical window spanning from November 30, 2020, to November 30, 2025, covering two years before and three years after Gen AI's release. We set the post-release cutoff at three years because sufficient data had accumulated stably by that point. We set the pre-release window at two years to allow for comparison with the pre-release period; we do not extend this window further back, however, because programming languages and software ecosystems evolve rapidly, and a longer historical window could undermine the consistency of our analysis. Among the various components of the Stack Overflow dataset, we focus on question posts, as they provide a detailed, time-stamped record of the technical knowledge users voluntarily seek and discuss (Fig.\ref{SI_dataset_B_1}).

For the first set of analyses, we build our primary dataset around Python-related questions, identified as posts explicitly tagged with \texttt{<python>} in the Stack Overflow dataset, yielding 662,894 questions for analysis. Python's ease of use, extensive open-source ecosystem, and versatility have driven its rapid adoption since the 2010s; it now serves as the lingua franca of artificial intelligence, machine learning, and beyond. We confirm this popularity in our own data: \texttt{<python>} is the most frequently used tag on Stack Overflow during our study period (Fig.\ref{SI_dataset_C_1}). Focusing on a single, widely used language also allows us to control for heterogeneity across programming languages, reducing potential confounding factors such as differences in syntax complexity, communication norms, and domain-specific usage patterns.

For the next analysis, we shift our analytical scope from a single-language focus to a cross-language perspective, extending our dataset to additional programming languages to test whether the patterns observed within Python generalize at this broader, aggregate level. We first extract all unique tags in our dataset, identifying 52,930 tags in total. These tags encompass various types, including programming languages (e.g., \texttt{<python>}, \texttt{<java>}), markup languages (e.g., \texttt{<html>}), frameworks (e.g., \texttt{<spring-boot>}, \texttt{<django>}), and technical concepts (e.g., \texttt{<multithreading>}, \texttt{<shared-memory>}), among others. From this set, we select tags corresponding to programming languages only, excluding frameworks and markup languages. This yields 30 distinct programming languages represented on Stack Overflow during our study period, spanning a wide range of maturity—from long-established languages such as Java, C\#, and C++ to more recently emerged ones such as TypeScript, Kotlin, and Dart. In total, this extended dataset comprises 2,272,334 questions across all 30 languages. Details of this process are provided in \siref{Supplementary Information}{sec:dataset_1.C}.

\subsection{Variable Extraction}

Motivated by the two hypotheses introduced above, we evaluate each question in our dataset along two dimensions: (1) difficulty, capturing whether a question demands high-level cognitive reasoning, and (2) data availability, capturing whether a question concerns a technical subject well represented in existing data. For each dimension, we use two operationalizations: one machine-based and the other human-based. Measuring the same concept through two independent operationalizations ensures the robustness of our findings. Table \ref{Data_Tbl1} provides an overview of these variables, with detailed explanations below.

\begin{table}[t!]
\centering
\caption{Variable Measurement}
\label{Data_Tbl1}
\renewcommand{\arraystretch}{1.25}

\begin{tabularx}{\textwidth}{
    p{0.16\textwidth}
    p{0.20\textwidth}
    >{\raggedright\arraybackslash}X
    p{0.20\textwidth}
}
\toprule
Variable & Measurement & Description & Question\newline component \\
\midrule

\textbf{Difficulty} 
& Question difficulty 
& Measures the technical difficulty of each question based on the rubric; evaluates technical sophistication.
& Natural language\newline Source code \\

& Code Complexity 
& Calculates cyclomatic complexity \citep{gill1991cyclomatic} in the source code; measures branching and decision points as a proxy for code complexity.
& Source code \\

\midrule

\textbf{Data\newline Availability} 
& Topic composition 
& Measures the composition of digitized content using topic modeling.
& Natural language \\

& Tag composition
& Measures the composition of digitized content based on user-assigned tags.
& Tag \\

\bottomrule
\end{tabularx}
\end{table}

To assess difficulty, we evaluate the content of each question, which includes both natural language (the question description) and programming language (the source code). Following a rubric developed for classifying Stack Overflow question difficulty \citep{raidaStudyClassifyingStack2024a}, four trained human annotators classify 324 randomly sampled questions into three difficulty levels: Basic, Intermediate, and Advanced. From these, we construct a gold-standard dataset of 124 questions for which all four annotators assign the same difficulty level. To scale this classification beyond what human annotation alone could cover, we develop an AI annotator using a quantized open-source LLM (Qwen3-30B-A3B-Instruct-2507) served via VLLM \citep{kwon2023efficient}, a framework-level library that accelerates LLM inference through the PagedAttention algorithm. 

Because applying the AI annotator to every question in our dataset would require substantial computational time, we instead randomly sample 30 Python-related questions per day across our entire 260-week study period, yielding a sample of 54,600 questions. This AI annotator labels these sampled questions, with its consistency against the human-annotated gold standard verified through systematic validation. As a robustness check, we also apply a different open-source LLM as the classifier and find patterns similar to those in Fig.\ref{C_Result_Fig1}A (see \siref{Supplementary Information}{sec:FE_2.A.7}). Detailed procedures for the annotation and data augmentation processes are provided in \siref{Supplementary Information}{sec:FE_2.A}.

As a second, complementary measure of difficulty, we employ a machine-based approach: cyclomatic complexity, applied to the source code embedded in each question. Cyclomatic complexity measures the number of independent decision points (e.g., conditional branches and loops) in a program, indicating how many distinct cases must be considered when understanding or testing the code \citep{mccabe1976complexity}. Although certain limitations of this metric have been noted \citep{shepperd1988critique}, its widespread adoption has produced reliable tool support across a broad range of programming languages \citep{gill1991cyclomatic}, making it well-suited for our multi-language analysis. Unlike the human-based difficulty measure, which relies on a sampled subset due to annotation cost, cyclomatic complexity can be computed directly from source code without such constraints. We therefore calculate a cyclomatic complexity score for every question containing source code in our dataset. Detailed procedures are provided in \siref{Supplementary Information}{sec:FE_2.B}.

For our second variable, data availability, we assess the volume of question posts accumulated on Stack Overflow for each technical subject. This measure is grounded in the ``scaling law''—the well-established finding that generative AI performs better when trained on larger corpora \citep{kaplan2020scaling}. Applied to our context, technical subjects with a greater accumulated post volume are more likely to be effectively handled by Gen AI, whereas rarely discussed subjects remain more dependent on human-generated knowledge. Because directly measuring the total volume of technical knowledge on the web is infeasible, we restrict our focus to Stack Overflow as a representative proxy for programming knowledge. Although imperfect, Stack Overflow served as the \textit{de facto} standard for software engineering knowledge exchange prior to Gen AI's release \citep{baltes2019usage, moutidis2021community}, making it a reasonable empirical site for assessing data availability.

We assess data availability using two complementary analytic units: a machine-assigned unit (topic) and a user-assigned unit (tag). To construct topic-based data availability, we extract latent topics from each question's text using BERTopic, identifying 50 distinct topics; for robustness, we also apply LDA to the same dataset and compare the results (see \siref{Supplementary Information}{sec:FE_2.C.2}). To construct tag-based data availability, we categorize questions using the user-assigned tags included in each Stack Overflow post. These two units carry different, complementary biases: user-generated tags reflect the author's intent but may also reflect strategic behavior unrelated to the question's content—for instance, authors adding popular tags to increase a post's visibility—whereas topic modeling is free from such human bias but may not fully capture the author's intended focus. Because of these differing biases, using both units allows us to test whether our results are sensitive to how data availability is operationalized. For detailed information on the topics and related procedures, see \siref{Supplementary Information}{sec:FE_2.C}.

\subsection{Analytic Strategies}   

Based on the measures extracted from our dataset, we analyze how software engineering knowledge has changed around Gen AI's release. Our analysis follows a four-step approach: first, we identify temporal changes in difficulty within Python-related questions; second, we apply the same analysis to data availability; third, we examine how these two dimensions jointly evolve; and finally, we test whether these patterns hold at a broader, cross-language level, beyond Python alone.

In the first step, we examine temporal changes in difficulty using two complementary measures. First, we calculate the weekly percentage distribution of questions across difficulty levels ($n = 260$, corresponding to the 260 weeks in our dataset) and apply a two-sided Chow test to detect statistically significant structural breaks, using an alpha level of 0.05. Second, we apply the same two-sided Chow test to the weekly average code complexity score ($n = 260$ weeks). Because the Chow test does not require strict normality assumptions, we do not conduct formal normality tests.

In the second step, we compare temporal trends in data availability before and after Gen AI's release. To capture how popular and rare technical subjects shifted over time, we first identify the top 20\% and bottom 20\% of topics and tags by frequency in the pre-Gen AI period. We then compute the weekly proportion of questions in each group ($n = 260$ weeks) and track how these proportions change across the pre- and post-Gen AI periods. To statistically test these temporal changes, we apply two-sided Chow tests to examine whether the slope of each group's weekly proportion changed around Gen AI's release, again using an alpha level of 0.05.

In the third step, we examine how difficulty and data availability jointly evolve over time. Examining these two dimensions together provides a more complete picture of how knowledge on Stack Overflow has been reshaped following Gen AI's release. For this analysis, we focus on user-assigned tags rather than machine-assigned topics, because topic modeling requires a predetermined number of topics and may therefore overlook newly emerging technical domains that do not fit existing clusters. User-assigned tags, in contrast, offer a more flexible classification of technical subjects without this constraint.

Drawing on the difficulty measure from the first step and the tag-based classification from the second step, we examine how the proportion of questions at each difficulty level differs between data-rich and data-scarce tags. To do so, we categorize tags into two groups based on their question volume in the pre-Gen AI period: data-rich tags, defined as the top 20\% by volume, and data-scarce tags, defined as the bottom 20\%. Within each group, we calculate the weekly proportion of questions across difficulty levels ($n = 260$ weeks) and estimate a segmented regression model within an ITS design \citep{bernal2017interrupted}. This model allows us to test, for each difficulty level and tag group, whether the weekly trend shifts significantly after Gen AI's release, while isolating this post-release change from any trend already present beforehand (see \siref{Supplementary Information}{sec:FE_3.A.2}). To further illustrate these patterns, we construct an ego-centric tag network centered on one illustrative tag (\texttt{openai-api}), in which nodes represent tags and edges connect tags that co-occur within the same questions. We then examine how this network differs across difficulty levels between the pre- and post-Gen AI periods (see \siref{Supplementary Information}{sec:FE_3.A.3}).
    
In the final step, we extend our analysis beyond Python to examine whether the observed patterns generalize across programming languages. For this cross-language comparison, we narrow our focus to two of the four measures—cyclomatic complexity and tag composition—because both are scalable and preserve consistency across languages. We select cyclomatic complexity because it relies on explicit, language-agnostic rules that require no additional specification, making it readily applicable across a broad range of languages; by contrast, constructing a consistent difficulty ruleset applicable across all 30 languages was not feasible for the human-based difficulty measure. However, computing weekly average complexity requires a sufficient number of code-containing questions in each week, so we restrict this measure to the 13 languages with adequate weekly volume of minimum 30 code snippets, excluding the remaining 17. Tag composition, by contrast, requires no such weekly volume threshold and is therefore computed across all 30 languages. Similarly, we select tag composition over machine-assigned topics because topic modeling requires optimizing the number of clusters separately for each language, which complicates cross-language comparison, whereas user-assigned tags require no such adjustment and thus allow direct comparison across languages.

To systematically compare code complexity and tag composition across programming languages, we divide this step into two stages. First, we examine overall trends across all languages to assess whether the aggregate patterns observed for Python hold more broadly. Second, we conduct language-level analyses to test whether the magnitude of these shifts varies across languages. For both stages, we employ the same ITS model \citep{bernal2017interrupted}, focusing on slope changes rather than intercept shifts, as the impact of Gen AI likely penetrates gradually rather than abruptly (see \siref{Supplementary Information}{sec:FE_3.B.2}).




\section*{Data Availability}
Stack Overflow data were obtained as a dump file from the publicly available site \url{https://archive.org/details/stackexchange}.

\section*{Code Availability}
The source code for reproducing the findings of this study is provided on GitHub repository \url{https://github.com/myokyunghan/uneven_automation.git}.

\section*{Supplementary Information}
Please contact the corresponding authors if you need the supplementary information file.

\clearpage 
\bibliographystyle{sn-mathphys-num}
\bibliography{sn-bibliography}






\end{document}